\documentclass{article}
\newcommand{\bb}{\begin{eqnarray*}}
\newcommand{\ee}{\end{eqnarray*}}
\begin{document}
\title{\Large {\bf Strong CP: No Problem}}
\author{P. Mitra\\
Saha Institute of Nuclear Physics\\
Calcutta 700064, India\\
parthasarathi.mitra {\it at} saha.ac.in}
\date{hep-ph/0504053}
\maketitle
\begin{abstract}
Detailed analysis shows that
the phase of a complex mass term of a quark does not violate CP, while
the QCD vacuum angle can naturally be set equal to zero. There is no
strong CP problem and no need for axions or similar speculative 
constructions to be experimentally looked for.
\end{abstract}
\vfill
\centerline{Talk given at THEP-I, IIT Roorkee, March 2005}
\vfill\newpage
\section*{Introduction}
Strong interactions basically differ from weak
interactions: vector
currents are involved instead of chiral currents and parity is
not essentially violated. However, one 
must look at other terms in the Lagrangian:
\bb
{\cal L}=\bar\psi(i\gamma^\mu D_\mu-me^{i{\theta'}\gamma_5})\psi
-\frac{1}{4}~{\rm tr}~F_{\mu\nu}F^{\mu\nu}
-n_f{g^2{\theta}\over
32\pi^2}~{\rm tr}~F_{\mu\nu}\tilde F^{\mu\nu}.
\ee
The topological term with the
QCD vacuum angle $\theta$ violates P and CP.
The quark mass term has a chiral phase
$\theta'$ from symmetry breaking in the
electroweak sector (it may be large $\approx 1$).
This may violate P and CP.
Why then is there no experimental observation?
This was referred to as the {\bf strong CP problem} \cite{scp}.
One may distinguish between the {\it first} strong CP problem -- 
why there is no observable effect of $\theta$ -- and the
{\it second} strong CP problem -- why there is no observable effect of $\theta'$
combined with $\theta$.
Under certain assumptions the two phases become interconvertible.
So some people define an effective parameter $\bar\theta\equiv\theta-\theta'$.
One then asks: is  $\theta-\theta'$=0?
Several attempts have been made to prove that $\theta$ 
does not really lead to CP violation in QCD \cite{tokarev,sachs,lee}.
In any case, the
first Strong CP problem can be solved by assuming that $\theta$=0.
Although zero is a  special value, this choice increases the symmetry of 
the action
and is {\it natural} according to 't Hooft's criterion of naturalness.
In contrast, making $\theta=\theta'$ is {\it unnatural}:
the symmetry of an effective action increases, but not of the classical action.
So the second strong CP problem cannot be solved this way.

A theoretical estimate of the CP-violating electric dipole moment of neutron 
is about $10^{-16}\bar\theta$
e-cm, to be compared with the experimental upper bound of $10^{-26}$ e-cm,
implying that $|\bar\theta|<10^{-10}$, 
requiring a high degree of {fine}-tuning.
Modifications of QCD have been proposed to avoid this need for fine tuning.
One approach is to use chiral symmetry.
A massless fermion makes $\theta$ unphysical. The
mass may be generated by Yukawa coupling to a complex scalar field
preserving the chiral symmetry of the
action and giving this field a vacuum expectation value.
There is then a Goldstone boson, the so-called {\bf axion}, which remains 
undetected in spite of extensive searches.
Another approach is to use symmetries to make  ${\theta'=0}$, so that
the second strong CP problem reduces to the first strong CP problem,
which can be solved as above by invoking naturalness. These models too 
have special features which have not found experimental support.
On the other hand, our group \cite{bcm} has shown that {$\theta'$}
{\bf does not in the final analysis lead to CP violation}.
Hence the second strong CP problem reduces to the first strong CP problem,
which can be solved as above by invoking naturalness.

Let us first look at the argument showing $\theta'$ to be CP-violating.
Mass terms are of the form
$\bar q_L {M} q_R+\bar{\tilde q_L}{\tilde M}\tilde q_R$+hc,
with complex matrices coming from symmetry breaking ($q,\tilde q$
refer to quarks with charges 2/3 and -1/3 respectively).
On diagonalization of ${M}, {\tilde M}$, by suitable transformations
\bb
q_L\rightarrow A_L^{-1}q_L,&&q_R\rightarrow A_R^{-1}q_R,\\
\tilde q_L\rightarrow  \tilde A_L^{-1}\tilde q_L,
&&\tilde q_R\rightarrow  \tilde A_R^{-1}\tilde q_R,
\ee
W-interactions pick up the matrix $A_L \tilde A_L^{-1}\equiv {C}$, 
the Cabibbo-\-Kobayashi-\-Mas\-kawa matrix, 
which may be complex, leading to CP violation in the {\bf weak} interactions. 
The diagonalized mass terms may be complex if $U(1)$ transformations are 
not used, leaving a scope for CP violation in the {\bf strong} interactions. 
\bb
\bar\psi_L me^{i\theta'}\psi_R+hc&=&\bar\psi me^{i\theta'\gamma_5}\psi\\
&=&\cos\theta'\bar\psi m\psi
+i\sin\theta'\bar\psi m\gamma_5\psi:
\ee
this looks like a scalar + a pseudoscalar,
suggesting parity violation.  
Furthermore, the phase factor $e^{i\theta'}\rightarrow e^{-i\theta'}$   under
an antilinear operation,
suggesting time-reversal violation.

A chiral transformation can however be used to remove $\theta'$:
$$\psi\rightarrow e^{-i\theta'\gamma_5/2}\psi,$$
$$\bar\psi\rightarrow \bar\psi e^{-i\theta'\gamma_5/2}.$$
$\theta'$ gets removed from the mass, but may reappear elsewhere:
$$\theta{g^2}F\tilde F/32\pi^2\rightarrow
(\theta-\theta'){g^2}F\tilde F/32\pi^2.$$
This is why some people talk of an effective parameter $\theta-\theta'$ 
for P, T violation.
But this {\bf anomaly} effect occurs at the one loop level
and is different from the tree level P, T violation
indicated by the preceding argument above. This should suggest caution.

An alternative way to remove $\theta'$ is through a  redefinition of 
the $\gamma$-matrices.
Since $\bar\psi=\psi^\dagger\gamma^0$,
\bb
{\cal L}&=&\bar\psi(i\gamma^\mu D_\mu-me^{i\theta'\gamma_5})\psi\\
&=&\psi^\dagger(i\gamma^0\gamma^\mu D_\mu-
m\gamma^0e^{i\theta'\gamma_5})\psi\\
&=&\psi^\dagger(i\widetilde{\gamma^0}\widetilde{\gamma^\mu} 
D_\mu-m \widetilde{\gamma^0})\psi,
\ee
where $\widetilde{\gamma^\mu}\equiv e^{-i\theta'\gamma_5/2}{\gamma^\mu}
e^{i\theta'\gamma_5/2}$.
Note that
$\widetilde{\gamma^\mu}\widetilde{\gamma^\nu}=\gamma^\mu\gamma^\nu$,
implying that standard anticommutation relations are satisfied.
Furthermore,
\bb
\widetilde{\gamma^\mu}^\dagger&=&e^{-i\theta'\gamma_5/2}{\gamma^\mu}^\dagger
e^{i\theta'\gamma_5/2}\\
\phantom{\widetilde{\gamma^\mu}^\dagger}&=&\pm e^{-i\theta'\gamma_5/2}
{\gamma^\mu}e^{i\theta'\gamma_5/2}\\
\phantom{\widetilde{\gamma^\mu}^\dagger}&=&\pm \widetilde{\gamma^\mu},
\ee
implying the same hermiticity properties as the parent matrices.
However, reality properties differ: time reversal has therefore to be checked.

\section*{Construction of Parity and Time-reversal}
For a real mass term, the parity matrix $P$ is defined through
\bb
&{\cal P}\bar\psi(i\gamma^\mu D_\mu-m)\psi{\cal P}^{-1}
=&\\
&{\cal P}\psi^\dagger{\cal P}^{-1}\gamma^{0}(i\gamma^{\mu} {\cal P}D_\mu
{\cal P}^{-1}-m){\cal P}\psi{\cal P}^{-1}
&\\
&=\psi^\dagger(-\vec r)P^\dagger\gamma^{0}(i\gamma^{0} D_0(-\vec r)
&\\
&+i\gamma^{i}
(-)D_i(-\vec r) -m)P\psi(-\vec r)&\\
&\stackrel{?}{=}
\psi^\dagger(-\vec r)\gamma^0(i\gamma^\mu D_\mu(-\vec r)
-m)\psi(-r),&\\
\ee
requiring
\bb
&P^\dagger P=1,&\\
&-P^\dagger\gamma^{0}\gamma^{i}P=\gamma^0\gamma^i,&\\
&P^\dagger\gamma^{0}P=\gamma^0.&
\ee
The second equation yields
\bb
-P^\dagger\gamma^{0}PP^\dagger\gamma^{i}P&=&\gamma^0\gamma^i,\\
\ee
whence
\bb
-P^\dagger\gamma^{i}P&=&\gamma^i.
\ee
This is known to be satisfied by
$P=\gamma^0$, leading to the transformation properties
\bb
{\cal P}\bar\psi \psi{\cal P}^{-1}
&=&\bar\psi(-\vec r) \psi(-\vec r),\\
{\cal P}\bar\psi\gamma_5 \psi{\cal P}^{-1}
&=&-\bar\psi(-\vec r)\gamma_5 \psi(-\vec r).
\ee
For a complex mass term, the situation changes:
\bb
&{\cal P}\bar\psi(i\gamma^\mu D_\mu-me^{i{\theta'}\gamma_5})
\psi{\cal P}^{-1}
=&\\
&{\cal P}\psi^\dagger{\cal P}^{-1}\gamma^{0}(i\gamma^{\mu} {\cal P}D_\mu
{\cal P}^{-1}-me^{i\theta'\gamma_5}){\cal P}\psi{\cal P}^{-1}
&\\
&=\psi^\dagger(-\vec r)P^\dagger\gamma^{0}(i\gamma^{0} D_0(-\vec r)
&\\
&+i\gamma^{i}
(-)D_i(-\vec r) -me^{i\theta'\gamma_5})P\psi(-\vec r)&\\
&\stackrel{?}{=}
\psi^\dagger(-\vec r)\gamma^0(i\gamma^\mu D_\mu(-\vec r)
-me^{i\theta'\gamma_5})\psi(-\vec r),&
\ee
requiring
\bb
P^\dagger P&=&1,\\
-P^\dagger\gamma^{0}\gamma^{i}P&=&\gamma^0\gamma^i,\\
P^\dagger\gamma^{0}e^{i\theta'\gamma_5}P&=&\gamma^0e^{i\theta'\gamma_5}.
\ee
The second equation yields
\bb
-P^\dagger\gamma^{0}e^{i\theta'\gamma_5}\gamma^{i}e^{i\theta'\gamma_5}P
&=&\gamma^0e^{i\theta'\gamma_5}\gamma^ie^{i\theta'\gamma_5},
\ee
whence
\bb
-P^\dagger\gamma^{i}e^{i\theta'\gamma_5}P&=&\gamma^ie^{i\theta'\gamma_5}.
\ee
This is satisfied by 
$P=\widetilde\gamma^0=\gamma^0e^{i\theta'\gamma_5}$ 
leading to the transformation properties
\bb
{\cal P}\bar\psi e^{i\theta'\gamma_5}\psi{\cal P}^{-1}
&=&\bar\psi(-\vec r) e^{i\theta'\gamma_5}\psi(-\vec r),\\
{\cal P}\bar\psi\gamma_5 e^{i\theta'\gamma_5}\psi{\cal P}^{-1}
&=&-\bar\psi(-\vec r)\gamma_5 e^{i\theta'\gamma_5}\psi(-\vec r).
\ee
The time-reversal matrix $T$ is defined for a real mass by
\bb
&{\cal T}\bar\psi(i\gamma^\mu D_\mu-m)\psi{\cal T}^{-1}
=&\\
&{\cal T}\psi^\dagger{\cal T}^{-1}\gamma^{0*}(-i\gamma^{\mu*} {\cal T}D_\mu
{\cal T}^{-1}-m){\cal T}\psi{\cal T}^{-1}
&\\
&=\psi^\dagger(-t)T^\dagger\gamma^{0*}(-i\gamma^{0*} (-)D_0(-t)
&\\
&-i\gamma^{i*}
D_i(-t) -m)T\psi(-t)&\\
&\stackrel{?}{=}
\psi^\dagger(-t)\gamma^0(i\gamma^\mu D_\mu(-t)
-m)\psi(-t),&
\ee
requiring
\bb
T^\dagger T&=&1,\\
-T^\dagger\gamma^{0*}\gamma^{i*}T&=&\gamma^0\gamma^i,\\
T^\dagger\gamma^{0*}T&=&\gamma^0.
\ee
The second equation yields
\bb
-T^\dagger\gamma^{0*}TT^\dagger\gamma^{i*}T&=&\gamma^0\gamma^i,
\ee
whence
\bb
-T^\dagger\gamma^{i*}T&=&\gamma^i.
\ee
In the standard representation of $\gamma$-matrices,
$\gamma^2$ is purely imaginary, while the rest are real. Then, 
\bb
T^\dagger\gamma^{\mu}T&=&\gamma^\mu (\mu=0,2),\\
T^\dagger\gamma^{\mu}T&=&-\gamma^\mu (\mu=1,3).
\ee
These are known to be satisfied by
$T=i\gamma^1\gamma^3$.

For a complex mass term, the situation changes:
\bb
&{\cal T}\bar\psi(i\gamma^\mu D_\mu-me^{i\theta'\gamma_5})\psi{\cal T}^{-1}
=&\\
&{\cal T}\psi^\dagger{\cal T}^{-1}\gamma^{0*}(-i\gamma^{\mu*} {\cal T}D_\mu
{\cal T}^{-1}-me^{-i\theta'\gamma_5^*}){\cal T}\psi{\cal T}^{-1}
&\\
&=\psi^\dagger(-t)T^\dagger\gamma^{0*}(-i\gamma^{0*} (-)D_0(-t)
&\\
&-i\gamma^{i*}
D_i(-t) -me^{-i\theta'\gamma_5^*})T\psi(-t)&\\
&\stackrel{?}{=}
\psi^\dagger(-t)\gamma^0(i\gamma^\mu D_\mu(-t)
-me^{i\theta'\gamma_5})\psi(-t),&
\ee
requiring
\bb
T^\dagger T&=&1,\\
-T^\dagger\gamma^{0*}\gamma^{i*}T&=&\gamma^0\gamma^i,\\
T^\dagger\gamma^{0*}e^{-i\theta'\gamma_5^*}T&=&\gamma^0e^{i\theta'\gamma_5}.
\ee
The second equation yields
\bb
-T^\dagger\gamma^{0*}e^{-i\theta'\gamma_5^*}\gamma^{i*}e^{-i\theta'\gamma_5^*}T&
=&\gamma^0e^{i\theta'\gamma_5}\gamma^ie^{i\theta'\gamma_5},
\ee
whence
\bb
-T^\dagger\gamma^{i*}e^{-i\theta'\gamma_5^*}T&=&\gamma^ie^{i\theta'\gamma_5}.
\ee

If ${\gamma}^\mu$ is in the standard representation, these are satisfied by 
$$T=e^{i\theta'\gamma_5}T_{standard}=ie^{i\theta'\gamma_5}\gamma^1\gamma^3.$$ 
\section*{The measure of functional integration}
We have seen that there are parity and time-reversal symmetries 
in the presence of
a complex mass term. The question is whether such symmetries
are broken by anomalies. Anomalies are often studied in a Euclidean framework:
\bb
S=\int \bar\psi(\gamma^\mu D_\mu-m)\psi.
\ee
With eigenfunctions $\phi_n$ of $i\gamma^\mu D_\mu$, one expands
\bb
\psi=\sum_n a_n\phi_n,\quad \bar\psi=\sum_n \bar{a}_n\phi^\dagger_n,
\ee
and constructs the functional integral
\bb
Z=\int {\cal D}A \prod_n \int da_n \prod_n \int d\bar{a}_n e^{-S}.
\ee
For a chiral transformation
\bb
\psi\rightarrow e^{i\alpha\gamma^5}\psi,\quad
\bar\psi\rightarrow\bar\psi e^{i\alpha\gamma^5},
\ee
the expansion coefficients transform as
\bb
a_n&\rightarrow&\sum_m\int\phi_n^\dagger 
e^{i\alpha\gamma^5}\phi_m a_m,\nonumber\\
\bar{a}_n&\rightarrow&\sum_m\bar{a}_m\int\phi_m^\dagger 
e^{i\alpha\gamma^5}\phi_n,
\ee
and there is a nontrivial Jacobian calculable with a
regularization involving gauge invariant eigenvalues of $i\gamma^\mu
D_\mu$: $e^{i\alpha g^2 F\tilde F/16\pi^2}$.
This amounts to a breakdown of chiral symmetry even if
$m=0$: this is the well known $U(1)$ anomaly.

For a complex mass term, the situation changes:
\bb
S=\int \bar\psi(\gamma^\mu D_\mu-me^{i{\theta'}\gamma^5})\psi,
\ee
\bb
\psi=e^{-i{\beta}\gamma^5/2}\sum_n 
a^\beta_n\phi_n,\quad \bar\psi=\sum_n 
\bar{a}^\beta_n\phi^\dagger_n e^{-i{\beta}\gamma^5/2},
\ee
\bb
Z=\int {\cal D}A \prod_n \int da^\beta_n \prod_n \int d\bar{a}^\beta_n 
e^{-S},
\ee
with a measure depending on a parameter $\beta$.

$\theta',\beta$ can be removed by chiral transformations  
at the expense of changes in $\theta F\tilde F$ term.
The effective parity violation parameter is seen to be
\bb
\bar\theta=\theta-\theta'+{\beta}.
\ee
Under a chiral transformation,
\bb
a^\beta_n&\rightarrow&\sum_m\int\phi_n^\dagger 
e^{i\alpha\gamma^5}\phi_m a^\beta_m,\nonumber\\
\bar{a}^\beta_n&\rightarrow&\sum_m\bar{a}^\beta_m\int\phi_m^\dagger 
e^{i\alpha\gamma^5}\phi_n.
\ee
This is unchanged ($\beta$-independent) and produces the standard anomaly.

Now under the parity operation for gauge fields,
\bb
\phi_n(x_0,\vec x)&\rightarrow&\gamma^0\phi_n(x_0,-\vec x),\\
\phi_n^\dagger(x_0,\vec x)&\rightarrow&\phi_n^\dagger(x_0,-\vec x)\gamma^0,
\ee
so that
\bb
\phi_n^\dagger(x_0,\vec x)e^{-i\beta\gamma^5/2}&\rightarrow&
\phi_n^\dagger(x_0,-\vec x)e^{-i\beta\gamma^5/2}
e^{i\beta\gamma^5}\gamma^0,\nonumber\\
e^{-i\beta\gamma^5/2}\phi_n(x_0,\vec x)&\rightarrow&\gamma^0
e^{i\beta\gamma^5}e^{-i\beta\gamma^5/2}\phi_n(x_0,-\vec x).
\ee
Comparing this with the parity transformation for the fermion, we see that
consistency (unchanged $a^\beta_n,\bar{a}^\beta_n$) can be achieved with 
\bb
\beta&=&\theta',\\ 
\bar\theta&=&\theta -\theta'+\beta=\theta.
\ee
This parity (and time-reversal) invariant
choice of the measure may be compared with the na\"{\i}ve measure:
\begin{center}
\begin{tabular}{|c|c|}
\hline
 {\bf Measure with} $\beta={0}$ & 
{\bf Measure with} $\beta={\theta'}$  \\
\hline
\hline
 {Agrees} with standard &  {Reduces} to standard \\
 measure for real  &  measure for real \\
 mass term &  mass term \\
\hline
 {Anomaly} standard &  {Anomaly} standard   \\
\hline
 Measure {symmetric} & Measure {symmetric} \\
 under {usual} & under  P, T  \\
 P, T but {not} & {symmetries of action} \\
 under symmetries &  but not under  \\
 of action &  usual P, T  \\
\hline
 &   \\
 ${\bar\theta=\theta-\theta'}$ &  {$\bar\theta=\theta$}\\
\hline
\end{tabular}
\end{center}
\section*{Regularization of action}
In an alternative approach, the generalized Pauli-Villars regularization,
the Lagrangian density is augmented to include some extra species:
\bb
{\cal L}_{\psi,~ reg}^{[0]}=
\bar\psi(\gamma^\mu D_\mu-m
)\psi+
\sum_j\sum_{k=1}^{|c_j|} \bar\chi_{jk}(\gamma^\mu D_\mu-M_j
)\chi_{jk}.
\ee
Here $\chi_{jk}$ are regulator spinor fields with 
fermionic or bosonic statistics, with
signs $\pm$ for integers $c_j$  satisfying
\bb
1+\sum_j c_j=0,\quad m^2+\sum_j c_jM_j^2=0,
\label{c}\ee
to cancel divergences.
Ultimately, one has to take the regulators $M_j\to\infty$. 

In the presence of a chiral phase $\theta'$, it is 
necessary to 
provide the same chiral phase in the regulator mass terms:
\bb
{\cal L}_{\psi,~ reg}^{[\theta']}=
\bar\psi(\gamma^\mu D_\mu-m
e^{i\theta'\gamma_5})\psi+
\sum_j\sum_{k=1}^{|c_j|} \bar\chi_{jk}(\gamma^\mu D_\mu-M_j
e^{i\theta'\gamma_5})\chi_{jk}.
\ee
This is then invariant under the redefined parity.

The measure of integration now includes the Pauli-Villars fields: 
\bb 
d\mu=d\psi d\bar\psi\prod_{jk} d\chi_{jk} d\bar\chi_{jk},
\ee
\bb
Z_{reg}^{[\theta']}\equiv\int d\mu e^{-\int d^4x 
{\cal L}_{\psi,~ reg}^{[\theta']}}.
\ee

Let us apply a chiral transformation to the physical fermion fields 
as well as to the regulators: there are
Jacobian factors corresponding to $\chi_{jk},\bar\chi_{jk}$
with powers $c_j/|c_j|$
because of fermionic/bosonic statistics, 
while $\psi,\bar\psi$ obey fermionic statistics:
\bb
J_{reg}=e^{i{(1+\sum_jc_j)}\int d^4x \alpha
g^2 (F\tilde F/16\pi^2)}=1.
\ee
In this regularized framework, the Jacobian 
for a {\it combined} chiral transformation on physical fermion
fields and regulators is thus trivial. Consequently,
\bb
Z_{reg}^{[\theta']}=Z_{reg}^{[0]},
\ee
showing $\theta'$ to be {\it unphysical}.

\section*{Lattice regularization} 
The Wilson regularization on the lattice reads
\bb
S=a^4\sum_x \bar\psi[\gamma^\mu {D_\mu+D^{*}_\mu\over 2}-
aD^{*}_\mu D_\mu -me^{i\theta'\gamma_5}]\psi,
\ee
where $a$ is the lattice spacing, $D_\mu$ is the forward covariant difference
operator on the lattice and $D^{*}_\mu$ is the  backward covariant difference
operator.  
The chiral anomaly manifests itself in the Wilson formulation
as an explicit breaking of the chiral symmetry of the action -- 
no non-trivial Jacobian arises from the measure.
The existence of the anomaly means that there is {\it no} regularization 
that can preserve the chiral symmetry.

The fermionic part of the action is {\it not} invariant under the
(redefined) parity, with the link variables transformed in the
usual way: does this signal a parity anomaly? The question is whether
{\it other} formulations of the lattice regularization can preserve the 
symmetry. 

One can introduce
a class of regularizations parametrized by $\theta''$:
\bb
S=a^4\sum_x \bar\psi[\gamma^\mu {D_\mu+D^{*}_\mu\over 2}
-aD^{*}_\mu D_\mu
e^{i\theta''\gamma_5} -me^{i\theta'\gamma_5}]\psi.
\ee
Such a
phase is also allowed in the theoretically popular actions satisfying 
the Ginsparg-Wilson relation $D\gamma_5+\gamma_5D=aD\gamma_5D$:
\bb
S=a^4\sum_x \bar\psi[e^{i\theta''\gamma_5\over 2}De^{i
\theta''\gamma_5\over 2} -me^{i\theta'\gamma_5}]\psi,
\ee
($D$= corresponding lattice Dirac operator).  

Suppose one chooses  $\theta''=\theta'$: 
the redefined parity, with the link variables transformed in the
usual way, leaves the fermionic action invariant!\footnote{The measure 
does not break 
symmetry even in the Ginsparg-Wilson case, as 
the chiral transformation involved here is a
normal chiral transformation, not involving $D$.}
This is a non-generic regularization, but the existence of {\it one} 
parity conserving
regularization means that the parity is not anomalous. 
If we take $a\to 0$ with $\theta''=\theta'$, the parity
continues to be conserved. The same is true of time-reversal.

\section*{{Conclusion}}

The parity and the time-reversal symmetries
are redefined in the presence of the chiral phase
$\theta'$ arising from the Higgs sector.  If the fermion measure
can be chosen, it has to have the symmetry of the fermion action, 
and then there is no parity or time-reversal violation in the 
strong interactions because of $\theta'$.
It is technically natural to adjust the value of $\theta$ to be zero 
in accord with the experimental results currently
negating parity and time-reversal
violation in the strong interactions. There is no non-trivial fine-tuning. 
Last but not least, this is in standard QCD and does not
require axions or similar hypothetical constructs loved by many speculators
and experimentalists.

\end{document}